# On the Quasi-Moment-Method as a Rain Attenuation Prediction Modeling Algorithm

Sulaiman Adeniyi Adekola, Ayotunde Abimbola Ayorinde, Hisham Abubakar Muhammed, Francis Olutunji Okewole, and Ike Mowete

*Abstract*— A computationally inexpensive, analytically simple, and remarkably efficient rain attenuation prediction algorithm is presented in this paper. The algorithm, here referred to as the Quasi-Moment-Method (QMM), has only two main requirements for its implementation. First, rain attenuation measurement data (terrestrial or slant path) for the site of interest must be available; and second, a model, known to have predicted attenuation for any site to a reasonable level of accuracy (base model), and whose analytical format can be expressed as a linear combination of its parameters, is also required. An important novelty introduced by the QMM algorithm is a normalization scheme, through which a modelling difficulty concerning exceedance probabilities outside a 0.01 to 1.00 range, is eliminated. Model validation and performance evaluation using a comprehensive set of data available from the literature clearly demonstrated that the QMM models consistently improved base model performance by more than 90%; and outperformed all published 'best fit' models with which they were compared.

*Index Terms*—ITU-R models, Normalized percentages of time, Quasi-Moment-Method, Rain attenuation prediction

## I. INTRODUCTION

RAIN ATTENUATION represents a key consideration in the design of wireless terrestrial and Earth-space radiocommunication links, whose performances degrade significantly on account of severe signal scattering and absorption associated with rainfall events, when operating frequencies are greater than 10GHz; [1] –[4]. Design procedures for these links, including signal fade mitigation strategies [5], [6], routinely utilize prediction models, typically developed with the use of field measurements. Easily the most popular of such models are the ITU-R models (exemplified by [7]- [9]), which are continuously being updated, and which have been put to extensive use by the global community of researchers, [4], [10]. The ITU-R models have however, been reported in the literature, to give inaccurate attenuation prediction outcomes, especially when utilized for rain attenuation in tropical climatic zones: underestimating in some cases, [11], [12]; over-estimating in some others [4], [13]; though yielding predictions comparable with measurements in a few other cases, [10], [14].

### A. Literature Review

Prediction inaccuracies and certain inconsistencies associated with the ITU-R models motivated the development of quite a few alternative models, [2]. These alternative models are generally classified as either 'statistical' (regression-based) or 'physical' (analytical), [1], [3]; though a recent review in [15] suggested that the five categories identified as 'empirical', 'physical', 'statistical', 'fade-slope', and 'optimization-based' may also apply, when the models are regarded from the formulation point of view. According to [1], the statistical models invariably rely on cumbersome regression algorithms, whereas the physical models, which provide more accurate prediction outcomes, are rather complex, requiring a large number of input parameters.

Quite a few of these models represent modifications of one ITU-R model or another, as for example, in the case of [16], in which the ITU-R P618-7 was modified in a best fit approach that utilized measurement data with MATLAB's 'lsqcurvefit'. The resultant model, which differed from the ITU-R model in the first two coefficients, was reported to have significantly improved prediction performance. In a slightly different approach described in [17] for slant path rain attenuation prediction, regression coefficients available in the literature and referred to as 'locally obtained' were adopted for use with the modification. Comparisons of predictions by the power law model that emerged with those due to a corresponding ITU-R model revealed that whereas the former performed better for frequencies lower

All the authors are with the Department of Electrical & Electronics Engineering of the University of Lagos, Akoka, Yaba, Lagos Nigeria.
emails and orcid id: adekoladeniyi43@gmail.com
https://orcid.org/0000-0002-3926-5378
aayorinde@unilag.edu.ng, https://orcid.org/0009-0005-9731-8924
hmuhammad@unilag.edu.ng, https://orcid.org/0009-0003-4312-6543
fokewole@unila.edu.ng, is https://orcid.org/0009-0009-5588-2358
amowete@unilag.edu.ng  https://orcid.org/0000-0002-8335-6170

than 20GHz, the reverse was the case for frequencies greater than 20GHz. Another best fit, regression-based approach was reported in [18], where rainfall rate and rain attenuation measurements supported the development of a non-linear regression algorithm. The process informed the determination of a conversion factor, through which average annual exceedance was related to average worst month exceedance in a manner similar to that prescribed by ITU-R P.841-6. Average worst month attenuation predicted by the model were compared with those due to P.837-1 and P.837-7 as well as a few other models, and outcomes reported suggesting that the P.837-7 model and that developed in [18] gave better results than other models; with the former performing best. Unlike the models discussed in the foregoing, the logarithmic and power-law best fit regression models developed in [12] for a 6.37km terrestrial link, did not derive directly from any existing model. Instead, the authors deduced analytical fits that best described the dependence of specific attenuation on rainfall rate. using measurement data for seven different rainy months. Best fit regression for rain attenuation prediction modeling as used in [3] focused on obtaining constants for a slant path adjustment factor model. Nonetheless, the algorithm also represents a modification of an ITU-R model in that in specifying 0.01% exceedance for an average year, factors that were functions of link elevation, polarization, and operating frequency are computed with the use of the equations of ITU-R P.838-3. Predictions by the model were significantly better than those due to all the other existing models with which it was compared. Investigations described by [19] utilized numerical data obtained from a volume integral equation formulation (VIEF) in the least-squares curve fitting (involving a combination of the method of differential corrections and Newton's iteration algorithm) formulation of the extinction cross section (ECS) of rain drops. An empirical formula for the determination of specific attenuation, which utilized that for the ECS, was then derived.

Learning Assisted Rain Attenuation (LARA) prediction models may be described as a class of modelling techniques, with basis in either Artificial Neural Networks (ANN) or Supervised Machine Learning (SML), for the model development process, [15]. An example of the ANN approach is provided by the contributions reported by [6], in which a Back Propagation NN (BPNN) model was developed with the use of measured rainfall data to predict rainfall rates; and thence, long term rain attenuation statistics. The model's ability to predict deep fades during storm events was rated 'sufficiently good', and its corresponding ability to support the implementation of dynamic fade mitigation strategies, adjudged 'satisfactory'. Similar outcomes were reported for the Feed Forward Backpropagation (FFB) Neural Network model developed in [20] for the prediction of rain attenuation along satellite links in South Africa. In this case, the FBB model's prediction performance was found to be comparable to that of the corresponding Simple Attenuation Model (SAM). These outcomes appear to support the justification offered in [21] for the use of an Evolution Programing Network (EPnet)-evolved ANN, as a preferable alternative to conventional ANN, for rain attenuation prediction modeling. For according to [21], conventional ANN leads to sub-optimum prediction models, whose error functions are susceptible to being trapped in local minima.

An excellent example of the use of the SML approach is available from [1], which described a regression-based algorithm that utilizes Gaussian Process (GP) compatible functions. By identifying how what the authors referred to as "a set of descriptive features" influence the desired outputs (assumed to have Gaussian distributions), the paper developed an approach to rain attenuation prediction, designed to reveal the nature of its dependence on such parameters as path length, operating frequency, polarization, and rainfall distribution profiles. A noteworthy feature of the model is its cross-application characteristics, through which the same model predicted rain attenuation for terrestrial links in different tropical countries with significant superiority over other models with which it was compared.

Rainfall drop size distribution (DSD) modeling is typically the first step in some semi-analytical techniques for rain attenuation prediction, with the contributions in [22] and [23] being representative examples. In the case of [22], two DSD models were proposed through the use of a 'method of moments' (involving third, fourth, and sixth moments) combined with lognormal and gamma distributions. With the DSD models, specific rain attenuation predictions models were prescribed through an integral expression, whose integrand is a product of extinction cross-section and the model for DSD. On the other hand, the DSD model utilized in [23] derived from a statistical evaluation of measurements, and was combined with Mie scattering (for the determination of extinction cross-section) to develop the integral expression modeling specific rain attenuation prediction. In addition to including an account for attenuation due to antenna wetness, the paper also reported the important conclusion that path reduction factor prescribed by ITU-R P.530-17 gives inaccurate outcomes when utilized for short-range links. This same conclusion featured in [24], in which the rain attenuation proposed for terrestrial line-of-sight links also includes an account for wet antenna effects and moderates inaccuracies due to the use of the conventional path adjustment factor with a rainfall adjustment factor.

The multi exponential cell (MultiEXCELL) simulation tool, described by [5] as a procedure for the generation of synthetic rain fields that are spatially correlated, has served as basis for many physical attenuation prediction models, [25], [26]. Its use in [25] facilitated the simulation of an interaction between a hypothetical terrestrial link and a synthetic rain field (described as realistic) through which the variation of path reduction factor with rain rate measured at the link's transmitter, operating frequency, and link path length, was investigated. Outcomes of the investigation informed the development of the analytical attenuation prediction model presented in the paper. A similar approach was utilized in [26] for the development of the publication's prediction model for free space optical links of length up to 5km. Although the model developed in [27] shares the similarity of a synthetic propagation environment with those of [25] and [26], it differs in that it utilized an atmospheric numerical simulator (ANS- consisting of a weather forecasting model and an electromagnetic module) for the computation of annual statistics of interest to rain attenuation prediction.

Another class of physical models are those referred as Synthetic Storm Techniques (SST) which, for Earth-space links, appear to have been pioneered by [28], where it was developed as a physical-mathematical approach to the modeling of the dynamics of rain attenuation. As originally conceived and utilized for rain attenuation prediction along slant paths as well as fade duration and fade rate of change (fade slope), the only physical input required (additional to operational parameters) is the 1-min rain rate time series at the location of interest. According to [28], the SST is suitable for an estimation of the storm translation speed, to enable the conversion of the rain rate time series to an equivalent space series along horizontal and slant paths. In a recent contribution concerning the use of the SST model for the prediction of rain attenuation in a hilly, heavy-rainfall tropical environment, [29], it was reported that the model over estimated both rain attenuation levels and fade slope: to suggest that it may require modification in order to perform accurately in tropical climates. It was pointed out in [30] that whereas SST combines the advantages of a solid physical-mathematical framework with the simplicity of requiring rain rate time series as only input, it suffers from a limitation arising from how it models the melting layer.

*B. Limitations of the Modeling Approaches*

A particularly useful summary of the main attributes of the modeling approaches briefly reviewed in the foregoing, is provided by [31]. According to that publication, the fundamental difference between the various rain attenuation prediction methods lies in how the time-space structure of rainfall rate is modelled. The paper posits that apart from the SST, which generates attenuation statistics through the conversion of rain rate / time distributions into rain rate / distance profiles, all other approaches essentially utilize local measurements of cumulative distributions rainfall rates. Some of these other modeling techniques specify the statistical profile of rain along the path of interest using one of two assumptions; either that of a single cell of suitable shape or that of several cells of a particular shape, with a prescribed statistical distribution of sizes. Still others introduce a reduction coefficient as a means of statistically profiling rainfall distribution; or in the alternative, adjust the actual path length, using a reduction factor to obtain an equivalent length, along whose extent, rainfall intensity may be regarded as uniform.

It has been pointed out in [2] and [31] that existing methods recommended by ITU-R for the prediction of rain attenuation along slant path and terrestrial links, all derive from the basic assumption of an equivalent cell in which rainfall rate is uniform as model for the rainfall rate's non-uniform distribution along the link. An associated observation by [2] and [31] is that prediction by these methods utilize only measured rainfall rate exceeded for 0.01% of the time with extrapolation techniques, to determine exceedance for other percentages of the time. These two characteristics, according to those two publications, are responsible for the main limitations (detailed in [2] and [31]) of the ITU-R methods. A semi-empirical method proposed by [2] addressed these limitations by retaining the equivalent cell concept, but modifying path adjustment factor modeling to achieve a consistency in the latter, when utilized for both slant path and terrestrial links.

*C. Synopsis of Paper Contents*

In this paper, a Quasi-Moment-Method (QMM) algorithm is introduced as a rain attenuation prediction modeling method, capable of modifying existing models to provide remarkably accurately performing models. For the purposes of validation and performance evaluation, measurement data from five different literature sources were utilized. Base models of varying complexity were taken as candidates for use with the development of the QMM models, whose performances not only remarkably improved the prediction accuracies of the base models, but were also better than the best fit models with which they were compared. An important novelty introduced in the paper is that of scaling "percentages of time" to range from 1% to 0.01%, and utilizing an 'equivalent $A_{0.01}$' in the QMM algorithm's implementation. This contribution is informed by a remark in [1], which, with reference to ITU-R P.837-7, noted that "*employing an empirical formula, the results obtained are scaled to percentages of time that range from 1% to 0.001%*".

Section II of this paper presents the analytical foundations of the QMM algorithm, as applied to rain attenuation prediction, whilst Section III discusses computational results concerning the validation of the algorithm and its performance evaluation. The main conclusions arising from the investigations as well as possibilities for extensions in future work are presented in Section IV, which is the paper's concluding section.

## II. PROPOSED APPROACH

*A. The Quasi-Moment-Method Algorithm*

Given a set of rain attenuation measurement data represented by $\{X(x_k)\}_{k=1}^{N}$ (where 'X' may denote either specific attenuation in dB/km or attenuation in dB; and $x_k$, either rain rate in mm/h or probability of exceedance in percentages), the prediction problem may be defined as that of determining the function $f(x)$, which is such that

$$f(x_k) = X(x_k) \quad \forall k. \quad (1)$$

The QMM algorithm specifies an approximation $f_a(x)$ to $f(x)$ in a manner that guarantees that the weighted Euclidean semi-norm of the error function

$$\varepsilon = f(x) - f_a(x) \quad (2)$$

assumes its smallest possible value: that is, [33], the numerical value of the quantity

$$\|\varepsilon\|^2 = \|f(x) - f_a(x)\|^2 = \sum_{k=1}^{N} |f(x_k) - f_a(x_k)|^2 w_k \quad (3)$$

in which $w_k$ is a weighting function, is the minimum possible. In particular, $f_a(x)$ is required to be derivable from a 'base' function $(f_b(x))$ known to have, with reasonable accuracy, predicted rain attenuation, and which will admit representation in the form [33], [34],

$$f_b(x) = \varphi_1(x) + \varphi_2(x) + \ldots + \varphi_M(x). \quad (4)$$

The set of functions $\{\varphi_k(x)\}$ appearing in (4) serve as basis and weighting functions as described for example, in [4]. And the problem then reduces to that of determining a set of coefficients $\{c_k\}$, such that

$$f_a(x) = c_1\varphi_1(x) + c_2\varphi_2(x) + \ldots + c_M\varphi_M(x), \quad (5)$$

satisfies (3).

These otherwise unknown coefficients are determined through a procedure, whose details are available in [33] – [35], to emerge in this case, as

$$\begin{pmatrix} c_1 \\ c_2 \\ \cdots \\ c_M \end{pmatrix} = \begin{bmatrix} \langle \varphi_1, \varphi_1 \rangle & \langle \varphi_1, \varphi_2 \rangle & \cdots & \langle \varphi_1, \varphi_M \rangle \\ \langle \varphi_2, \varphi_1 \rangle & \langle \varphi_2, \varphi_2 \rangle & \cdots & \langle \varphi_2, \varphi_M \rangle \\ \cdots & \cdots & \cdots & \cdots \\ \langle \varphi_M, \varphi_1 \rangle & \langle \varphi_M, \varphi_2 \rangle & \cdots & \langle \varphi_M, \varphi_M \rangle \end{bmatrix}^{-1} \begin{pmatrix} \langle \varphi_1, X \rangle \\ \langle \varphi_2, X \rangle \\ \cdots \\ \langle \varphi_M, X \rangle \end{pmatrix} \quad (6)$$

provided that the inner product quantities appearing in (6) are defined by

$$\langle \varphi_m, \varphi_n \rangle = \sum_{k=1}^{N} \varphi_m(x_k) \varphi_n(x_k), \quad (7a)$$

and

$$\langle \varphi_m, X \rangle = \sum_{k=1}^{N} \varphi_m(x_k) X(x_k). \quad (7b)$$

As pointed out in [33] and [35] and demonstrated in [34], the solution to the prediction problem given by (6) is unique when all the basis functions in the set $\{\varphi_k\}$ are linearly independent.

*B. Base Models for the Algorithm.*

Because they are arguably the most widely used rain attenuation prediction models by the global research community, the ITU-R models satisfy one of the key requirements for QMM base models, and consequently represent natural choices for $f_b$. However, these models, which are typically of the forms [7], [8]

$$\gamma = K R^\alpha \quad (8a)$$

and

$$A_p = A_{0.01} B_1 p^{-(B_2 + B_3 \log_{10} p)} \quad (8b)$$

for specific attenuation and probability of p% $(p \neq 0.01\%)$ exceedance for the average year, respectively, do not satisfy the requirement of (4). In this paper therefore, prior to the implementation of the QMM algorithm, base models in the forms of (8a) and (8b) are first recast into formats that satisfy the requirement specified by (4). Thus, taking the natural logarithm of both sides of (8a) yields

$$\log_e \gamma = \log_e K + \alpha \log_e R, \quad (9)$$

for which

$$f_a = c_1 \log_e K + c_2 (\alpha \log_e R). \quad (10)$$

The desired unknown coefficients are then determined by the QMM algorithm as

$$\begin{pmatrix} c_1 \\ c_2 \end{pmatrix} = \begin{bmatrix} \langle \log_e K, \log_e K \rangle & \langle \log_e K, \alpha \log_e R \rangle \\ \langle \alpha \log_e R, \log_e K \rangle & \langle \alpha \log_e R, \alpha \log_e R \rangle \end{bmatrix}^{-1} \begin{pmatrix} \langle \log_e K, \log_e \gamma \rangle \\ \langle \alpha \log_e R, \log_e \gamma \rangle \end{pmatrix} \quad (11)$$

The corresponding prediction model is thereafter obtained as

$$\gamma_{QMM} = exp(c_1 \log_e K + c_2 (\alpha \log_e R)) = K^{c_1} R^{\alpha c_2}. \quad (12)$$

In like manner, taking the natural logarithm of both sides of (8b) enables the definition of an associated base model according to

$$f_a = c_1 \log_e A_{0.01} + c_2 \log_e B_1 + c_3 (-B_2 \log_e p) + c_4 (-B_3 \log_{10} p (\log_e p)). \quad (13)$$

And following the definitions of the matrix

$$[\Phi] = \begin{bmatrix} \langle \varphi_1, \varphi_1 \rangle & \langle \varphi_1, \varphi_2 \rangle & \langle \varphi_1, \varphi_3 \rangle & \langle \varphi_1, \varphi_4 \rangle \\ \langle \varphi_2, \varphi_1 \rangle & \langle \varphi_2, \varphi_2 \rangle & \langle \varphi_2, \varphi_3 \rangle & \langle \varphi_2, \varphi_4 \rangle \\ \langle \varphi_3, \varphi_1 \rangle & \langle \varphi_3, \varphi_2 \rangle & \langle \varphi_3, \varphi_3 \rangle & \langle \varphi_3, \varphi_4 \rangle \\ \langle \varphi_4, \varphi_1 \rangle & \langle \varphi_4, \varphi_2 \rangle & \langle \varphi_4, \varphi_3 \rangle & \langle \varphi_4, \varphi_4 \rangle \end{bmatrix} \quad (14)$$

for which

$$\begin{Bmatrix} \varphi_1 \\ \varphi_2 \\ \varphi_3 \\ \varphi_4 \end{Bmatrix} = \begin{Bmatrix} \log_e A_{0.01} \\ \log_e B_1 \\ -B_2 \log_e p \\ -B_3 \log_{10} p (\log_e p) \end{Bmatrix}; \quad (14a)$$

as well as the column vector

$$(\Psi) = \begin{pmatrix} \langle \log_e A_{0.01}, X \rangle \\ \langle \log_e B_1, X \rangle \\ \langle -B_2 \log_e p, X \rangle \\ \langle -B_3 \log_{10} p (\log_e p), X \rangle \end{pmatrix}, \quad (15)$$

the solution to the problem (determination of the coefficients $c_k$) emerges in this case, as

$$\begin{pmatrix} c_1 \\ c_2 \\ c_3 \\ c_4 \end{pmatrix} = [\Phi]^{-1} (\Psi). \quad (16)$$

Hence, the corresponding QMM prediction model becomes given by

$$(A_p)_{QMM} = (A_{0.01})^{c_1} (B_1)^{c_2} p^{-(c_3 B_2 + c_4 B_3 \log_{10} p)} \quad (17)$$

Computational results presented in Section III derive in the main, from analytical models obtained with the use of the procedure outlined in the foregoing discussions. In each of the cases considered, measurement data were obtained from journal publications, through the use of the commercial graph digitizer software, GETDATA.

## III. VALIDATION AND PERFORMANCE EVALUATION

### A. Comparison with Results of [32]

As a first simple example, measurement data available from [32, Fig. 8] were utilized for the development of a QMM specific attenuation prediction model of the type given by (12). Predictions by the best fit model proposed by [32] as

$$\gamma_R = 0.567 R^{0.791} \tag{18}$$

are compared in Fig. 1 with those due to the corresponding QMM model obtained here as

$$\gamma_{QMM} = 1.05^{-18.8061} R^{0.77(1.1108)}. \tag{19}$$

In (19), the coefficients produced by the QMM algorithm are identified by magenta coloured fonts.

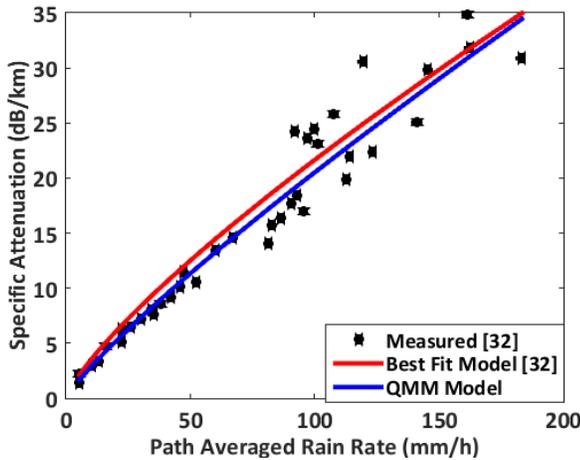

**Fig. 1.** Comparison of the performances of the best fit model of [32] with the corresponding QMM model.

The profiles of Fig. 1 clearly show that both models have comparable prediction performances; and this is confirmed by their Root Mean Square prediction Errors (RMSE), recorded as 2.4540 for [32]'s best fit model and 2.3148 for the QMM model.

### B. Performance Comparison with Best Fit Models

A demonstration of the veracity of the base model requirements, as described in Section II is offered by the following example, taken from [12, Section 5.2]. It concerns the prediction of rain attenuation along a 6.73km, 19.5GHz terrestrial link for six different months, using the best fit models listed in Table 4 of the publication. According to the information provided by the table, either a 'logarithmic' model or a 'power law' model (and not both) serves as best fit for any given of the six months considered. For the implementation of the QMM algorithm in this example therefore, identical base models were utilized for each of the six cases to obtain the QMM models given as

$$(\gamma)_{\lg r} = c_1(3.25 \log_e R) + c_2(2.500) \tag{20}$$

for the logarithmic base model, and

$$(\gamma)_{plaw} = \exp\left(c_1\left(\log_e(1.05)\right) + c_2\left(0.77 \log_e R\right)\right)$$
$$= 1.05^{c_1} R^{0.77 c_2}, \tag{21}$$

for the power law base model. It may be noted from (19) and (21) that the power law base model is the same as utilized in the first example. Numerical values for the model coefficients ($c_1$ and $c_2$) of (20) and (21) due to the implementation of the QMM algorithm, using measurement data available from those of [12, Fig. 7 -13], corresponding to the months listed in the paper's Table 4, are displayed in Table I below.

TABLE I
MODEL COEFFICIENTS FOR THE QMM MODELS OF (20) AND (21)

| Month / Model | Logarithmic Model | | Power Law Model | |
|---|---|---|---|---|
| | $c_1$ | $c_2$ | $c_1$ | $c_2$ |
| February | 1.2605 | 1.3924 | 30.1979 | 0.5915 |
| March | 1.5107 | -1.1246 | 0.1766 | 1.1447 |
| April | 0.6628 | 0.0956 | -6.3496 | 1.0875 |
| September | 0.6301 | 1.2035 | 23.7815 | 0.4245 |
| November | 3.7432 | -2.9339 | -5.3720 | 1.7486 |
| December | 1.0427 | -0.5471 | -0.9829 | 1.0205 |

A comparison of the performances of the QMM models defined by Table I and (20) and (21), with the corresponding best fit models of [12, Table 4] is provided by the prediction profiles of Fig. 2

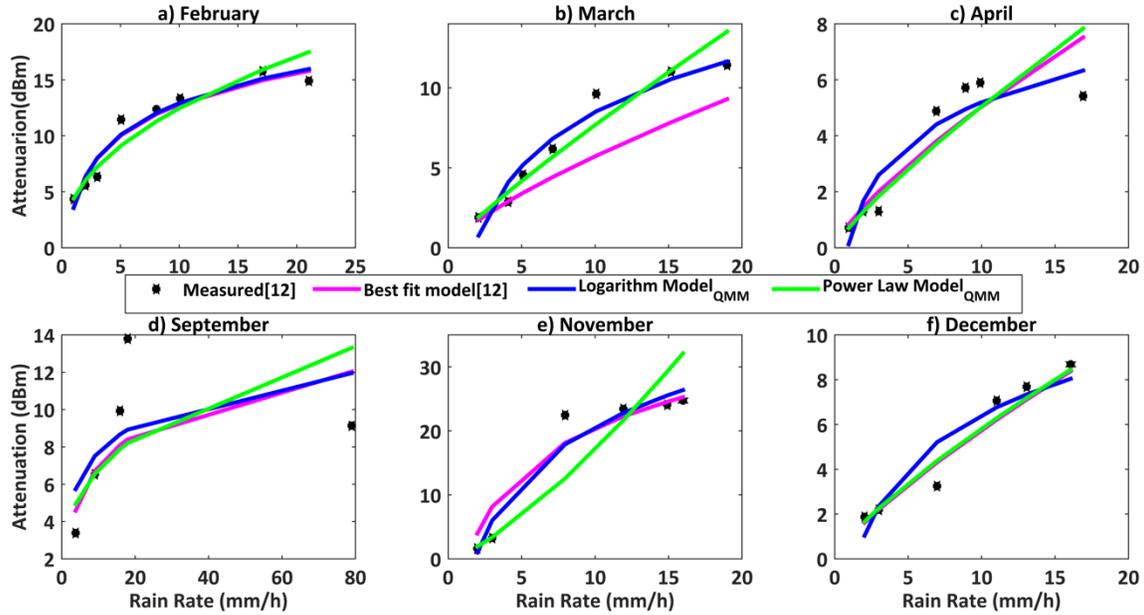

**Fig. 2.** Attenuation prediction profiles of the best fit models of [12] with the corresponding QMM models.

It is apparent from the curves of Fig. 2 that the QMM models are generally better performing than the best fit models defined by [12, Table 4].

TABLE II
RMSE PREDICTION METRICS FOR THE QMM AND BEST FIT MODELS OF FIG. 2

| Month / Model | Best Fit Model [12] | | QMM Models | |
|---|---|---|---|---|
| | *Power-Law* | *Logarithmic* | *Power-Law* | *Logarithmic* |
| February | N/A | 0.9868 | 1.3670 | 0.9813 |
| March | 2.2038 | N/A | 1.1130 | 0.8482 |
| April | 1.0758 | N/A | 1.1696 | 0.7841 |
| September | N/A | 2.9093 | 3.3255 | 2.8190 |
| November | N/A | 2.8876 | 5.5212 | 2.3846 |
| December | 0.6243 | N/A | 0.6118 | 0.9311 |

This observation is supported by the RMSE metrics of Table II, from which a number of inferences may be made. First, the models listed in [12, Table 4] clearly imply that for the months of March, April, and December, power-law models gave better best fit predictions than the logarithmic models; and that the reverse is the case for February, September, and November. The metrics of Table II suggest however, that when evaluated in terms of RMSE, the QMM-logarithmic models are generally best for prediction along this link. And second, with the exception of April (for the power law models), all the QMM prediction models recorded better RMSE values than corresponding best fit models of [12].

*C. Modeling with Measurement Results from [11]*

One of the main purposes served by the third example, subject of ensuing discussions, is to highlight an important outcome of the investigations reported in this paper. Computational results indicate that when implementing the QMM algorithm for the prediction of rain attenuation cumulative distribution function (CDF), significantly improved results are obtained, if exceedance percentages are scaled (or normalized) to ensure that the lower limit of the range of percentages is 0.01%. In that connection, and using measurement results from [11] for four different links, two QMM prediction models are developed, using the same base model of the type specified by 8(b), and given by

$$f_b = \log_e 0.12 + \log_e A_{0.01} - 0.60 \log_e p - 0.06(\log_{10} p)\log_e p. \quad (22)$$

The first QMM model (referred to as 'model 1') is easily shown to be given by

$$(A_{\%p})_{m1} = 0.12^{c_1}(A_{0.01}^n)^{c_2}(p_n)^{-(0.6c_3 + 0.06c_4 \log_{10}(p_n))} \quad (23)$$

provided that $A_{0.01}^n$ is the normalized rquivalent of $A_{0.001}$, and $p_n = 10p$,

Accordingly, the following equivalents apply;

$$0.001\% \leq p \leq 0.1\% \quad (24a)$$

corresponds to

$$0.01\% \leq p_n \leq 1\% \quad (24b)$$

The second QMM model (model 2), is defined by

$$\left(A_{\%p}\right)_{m2} = 0.12^{c_1} A_{0.01}^{c_2} (p)^{-\left(0.6 c_3 + 0.06 c_4 \log_{10}(p)\right)} \quad (25)$$

with the range for p specified by (24a), in this case.

For the computational results due to the models, measurement data were extracted from [11, Fig. 3]; $A_{0.01}$, from the paper's Table 2, and $A_{0.001}$, from the graphical presentation of its measurement data. Model parameters for the two QMM models then emerged as described by Tables III and IV, for models 1 and 2, respectively.

TABLE III
MODEL PARAMETERS FOR QMM MODEL 1

| Model Parameters /Frequency | 15GHz | 22GHz | 26GHz | 38GHz |
|---|---|---|---|---|
| $A_{0.01}^n$ | 8.9069 | 17.3337 | 23.3699 | 32.5993 |
| $c_1$ | -1.4788 | -0.6730 | 0.7774 | -1.0549 |
| $c_2$ | -1.1696 | -0.1089 | 0.9927 | -0.1332 |
| $c_3$ | 2.6014 | 2.9788 | 2.7041 | 2.8204 |
| $c_4$ | 16.9102 | 20.4688 | 17.5437 | 18.7999 |

TABLE IV
MODEL PARAMETERS FOR QMM MODEL 2

| Model Parameters /Frequency | 15GHz | 22GHz | 26GHz | 38GHz |
|---|---|---|---|---|
| $A_{0.01}$ | 4.2 | 8.3 | 10.98 | 15.10 |
| $c_1$ | -11.0996 | -0.0191 | -1.0720 | 0.3269 |
| $c_2$ | -18.2559 | -1.1863 | -1.7347 | -0.3584 |
| $c_3$ | 3.0434 | 3.3915 | 3.1671 | 3.2441 |
| $c_4$ | 6.2563 | 7.0388 | 6.5214 | 6.6857 |

A comparison of rain attenuation predicted by these models with corresponding measurements (as well as those due to the base models) for each of the four links is displayed in Fig. 3. Profiles in the figure very clearly reveal that both QMM models represent remarkably improved versions of the base model from which they derived. A close look at the profiles of QMM models 1 and 2 reveals that for 'percentages of time' greater than 0.01, predictions by both models are comparable; though with model 1 consistently giving better results. However, for values of p less than 0.01, model 2's performance is distinctly the poorer of the two.

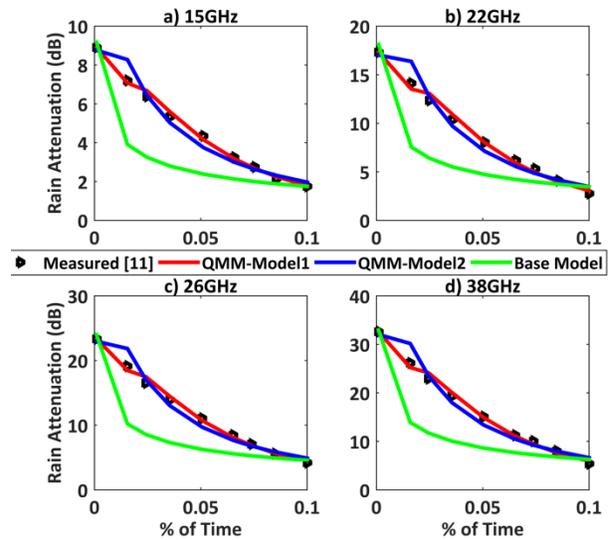

**Fig. 3.** Comparison of attenuation predictions by QMM and base models with measurements available from [11].

It would appear therefore that by arranging to ensure that attenuation for percentages of time other than 0.01% emerge, through extrapolation, from an 'equivalent' $A_{0.01}$, model 1 guarantees a better prediction performance. The RMSE metrics of Table V quantify the extent to which the base and QMM models track measurement data.

TABLE V
ROOT MEAN SQUARE PREDICTION ERROR METRICS FOR THE MODELS OF FIG. 3.

| Frequency (GHz) | QMM-Model1 RMSE(dB) | QMM-Model2 RMSE(dB) | Base Model RMSE(dB) |
|---|---|---|---|
| 15 | 0.1599 | 0.4441 | 1.9130 |
| 22 | 0.4057 | 0.9243 | 3.6593 |
| 26 | 0.4647 | 1.1425 | 4.9806 |
| 38 | 0.6394 | 1.6391 | 6.8952 |

It is readily verified from Table V that in terms of RMSE, model 1 represents an average of close to 60% performance improvement over model 2, and of about 91% compared with the base model.

*D. Model Calibration with Measurements from [1]*

Support for the conclusions described in the foregoing example is offered by performances of the QMM models developed from the use of measurement data available from [1], for which exceedance probability is in the range $10^{-5} \leq p \leq 10^{-3}$. With these values of p, it is not ordinarily possible to extrapolate from known values of $A_{0.01}$. And in order to negotiate the difficulty posed by that observation, an equivalent $A_{0.01}$ is defined in this paper through scaling (normalization) to set the range of p to $10^{-2} \leq p_n \leq 10^0$, in implementing the QMM algorithm. For the base model, use is made of the Chinese model given by [31, (10)], such that

$$\begin{Bmatrix} \varphi_1 \\ \varphi_2 \\ \varphi_3 \\ \varphi_4 \\ \varphi_5 \\ \varphi_6 \end{Bmatrix} = \begin{Bmatrix} \log_e (A_{0.01})_{eq} \\ -0.854 \log_e \left( \frac{p_n}{0.01} \right) \\ \left( 0.026 \frac{(\log_e (1+p_n))}{p_n} \right) \log_e \left( \frac{p_n}{0.01} \right) \\ \left( 0.022 \log_e ((A_{0.01})_{eq}) \right) \log_e \left( \frac{p_n}{0.01} \right) \\ (0.03 \log_e f) \log_e \left( \frac{p_n}{0.01} \right) \\ (0,226(1+p_n)) \log_e \left( \frac{p_n}{0.01} \right) \end{Bmatrix}, (26)$$

in which $f$ (GHz) represents operating frequency, with

$$p_n = 10^3 p; \quad (27a)$$

and

$$(A_{0.01})_{eq} = A_{0.00001}. \quad (27b)$$

The implementation of the associated QMM algorithm utilized measurement data from [1, Fig. 15, 16, 17, and 18]. Model parameters for the resulting models are displayed in Table VI below.

TABLE VI
QMM MODEL PARAMETERS FOR THE BASE MODEL DEFINED BY (26) AND (27)

| Parameters | Figure 15 f = 32,6GHz | Figure 16 f = 57GHz | Figure 17 f = 97GHz | Figure 18 f = 137GHz |
|---|---|---|---|---|
| $c_1$ | 0.9588 | 1.0857 | 1.0417 | 0.9050 |
| $c_2$ | -0.3013 | -0.0028 | -0.5665 | -5.1040 |
| $c_3$ | -0.7226 | -3.5261 | -2.5008 | 0.8328 |
| $c_4$ | 13.2114 | 9.8083 | 0.4972 | -28.8411 |
| $c_5$ | -8.0456 | -0.6046 | -1.4165 | -9.6709 |
| $c_6$ | -1.7394 | -1.9079 | -1.1351 | -2.1357 |
| $(A_{0.01})_{eq}$ | 47.9124 | 17.6559 | 46.0963 | 37.0937 |

Rain attenuation predictions due to the QMM models defined by the parameters of Table VI are compared in Fig. 4, with corresponding measurement data, as well as those due to the associated base models.

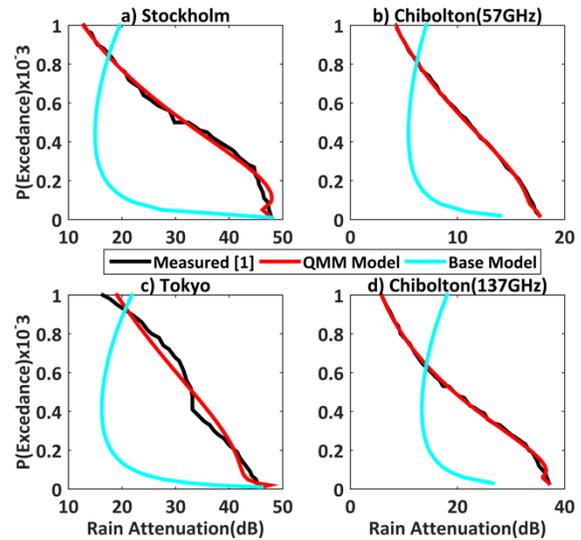

**Fig. 4.** Profiles of rain attenuation predicted by the QMM and base models defined by (27) and Table VI, compared with corresponding measured attenuation profiles from [1].

RMSE metrics, which characterize the performances of the models described by Fig. 4 were obtained as (QMM$_{RMSE}$, Base$_{RMSE}$) = (0.8990, 17.9714), (0.1169, 5.8819), (1.4958, 15.7974), and (0.4326, 12.1586) for the profiles of Fig. 4(a), 4(b), 4(c), and 4(d), respectively. The consistently excellent prediction profiles of the QMM models lend further credence to the efficacy of the scaling procedure introduced by this paper.

*E. Comparison with Measurements from [3]*

A final example considered in this paper developed QMM models with the use of three sets of data extracted from [3]. For base model in this case, the model proposed by [3, (4)] was adopted, to give

$$\begin{Bmatrix} \varphi_1 \\ \varphi_2 \\ \varphi_3 \\ \varphi_4 \\ \varphi_5 \end{Bmatrix} = \begin{Bmatrix} \log_e A_{0.01} \\ -1.0063 \log_e (p/0.01) \\ (-0.0591 \log_e p) \log_e (p/0.01) \\ (0.1317 \log_e A_{0.01}) \log_e (p/0.01) \\ (\beta(1-p) \sin \vartheta) \log_e (p/0.01) \end{Bmatrix} \quad (28)$$

Outcomes of the use of this base model for the development of QMM prediction models are presented in Table VII below.

TABLE VII
PARAMETERS OF QMM MODELS DERIVED FROM THE BASE MODEL OF (28)

| Parameters | Cameroon f=11.6GHz $\vartheta=47^0$ | Nigeria f =11.6GHz $\vartheta=48.3^0$ | Eindhoven f = 28,7GHz $\vartheta=26.9^0$ |
|---|---|---|---|
| $c_1$ | 1.1346 | 1.0176 | 0.9122 |
| $c_2$ | 0.1147 | 0.8082 | 0.3930 |
| $c_3$ | -1.5232 | 0.1693 | 1.5566 |
| $c_4$ | -1.2625 | 0.6891 | 0.1957 |
| $c_5$ | 16.8279 | 9.8414 | 8.6792 |
| $A_{0.01}$ | 25.4999 | 23.0226 | 23.3953 |
| $\beta \sin \vartheta$ | 0.0381 | 0.0426 | 0.0226 |

And the profiles of Fig. 5 describe how predictions by the QMM models and their associated base models compare with measurement data, for the three sets of data considered.

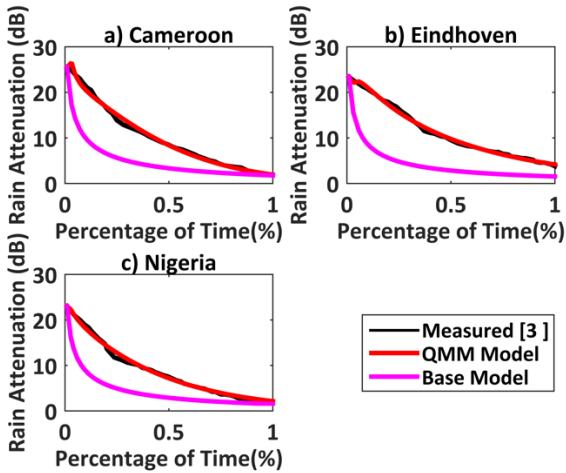

**Fig. 5.** Comparison of rain attenuation predicted by QMM models developed with (8) and data from [3], with the measurement data.

With root mean square prediction errors recorded as 0.5713dB, 0.3555dB, and 0.4138dB, for the QMM models developed using the 'Cameroon', 'Eindhoven', and 'Nigeria' measurement data, respectively, computations reveal that on the average, QMM modeling produced close to a 93% improvement on the prediction performances of the base models. It is worth pointing out that this example also serves to underscore the fact that no scaling (or normalization) scheme is required for the implementation of the QMM algorithm, when exceedance probability is such that $0.01 \leq p \leq 1.0$.

IV. CONCLUDING REMARKS

This paper has presented the Quasi-Moment-Method (QMM) as a tool for the modeling of rain attenuation prediction along both slant path and terrestrial links. A succinct, but representative presentation of the algorithm for QMM rain attenuation modeling identified the two important requirements of 'base models' as i) being known to have predicted rain attenuation anywhere, with reasonable accuracy; and ii) having an analytical format that will admit expression as a linear combination of the model parameters. Using a comprehensive set of measurement data extracted from five different publications in the literature, the validity of the modeling algorithm was established with examples of varying complexities. In addition, and through the same examples, the performances of various QMM models were evaluated, and found to be excellent. For example, compared against a number of published best fit models, the QMM models recorded root mean square prediction errors (RMSE) that were an average of close to 60% better. As a matter of fact, the RMSE values recorded for the QMM models ranged between 0.1599dB and 2.9dB (excluding those for 'power law models of Table II), indicative of a remarkably excellent performance. Indeed, these RMSE values were, in many cases, consistently greater than 90% better than those of the base models from which they derived.

One other main contribution of the paper is the establishment, through two examples (involving eight (8) different data sets) of the fact that when the range of exceedance probabilities (p) specified by measurement data is outside $0.01\% \leq p \leq 1\%$, the normalization scheme developed as part of the QMM modeling algorithm, leads to near perfect prediction models.

Finally, outcomes of investigations reported in the paper indicate a number of possibilities for future work. The more obvious of these may be set forth in the form of the following questions:

a) are there existing models, which satisfy the requirements of the QMM algorithm, and through which the inner product quantities on both sides of (6) can be given useful physical interpretations concerning electromagnetic wave-matter interaction ?
b) is it feasible to hybridize the QMM algorithm with some machine learning-based or physical rain attenuation prediction models?
c) how can the QMM algorithm be implemented towards developing a 'cross-application' model of the type proposed in [1]?

On-going investigations are directed at providing possible answers to these questions.

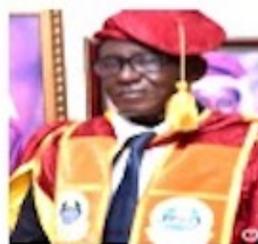

**Sulaiman Adeniyi Adekola** was awarded the degrees of B.Eng. (Hons.) Electrical Engineering at the Ahmadu Bello University (ABU) Zaria, Nigeria, 1968; M. Sc. Electrical Engineering at the Ohio State University, Columbus, Ohio, USA, 1972; Ph.D. in Engineering at the Ohio State University, Columbus, Ohio, USA, 1975.His research interests extend over Engineering Electromagnetics, Antenna Analysis & Design, Applied mathematics, Communications and Digital Signal Processing. An Emeritus Professor, Electrical & Electronics Engineering, UNILAG, 2019; Adekola is a Fellow, British Institution of Engineering & Technology (FIET), 1988; Fellow, Acoustical Society of America (FASA), 1988, Fellow, Nigerian Society of Engineers (FNSE), 1994; Fellow, Nigerian Academy of Science (FAS), 1990; Foundation Fellow, Nigerian Academic of Engineering (FAEng.), 1999; Life Snr Member, IEEE, USA, 2011; and a Chartered Engineer (C. Eng.).

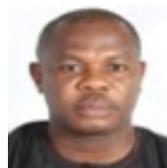

**Ayotunde Ayorinde** received the B.Sc., M. Sc. and Ph.D. degrees in Electrical Engineering, from the University of Lagos, (UNILAG) Nigeria, in 1990, 1993, and 2009, respectively. His current research interests include electromagnetic fields, microwave engineering, antennas and propagation. He is one of the pioneering investigators of the use of the QMM in empirical modelling of interest to radiowave propagation. Dr. Ayorinde, a Nigerian Registered Professional Engineer, is a Senior Lecturer at the University of Lagos, and is also a member of the British Institution of Engineering and Technology (MIET).

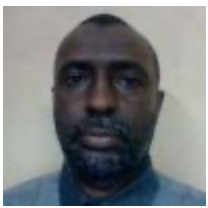

**Hisham Muhammed** received the B. Eng. and M. Sc. degrees in Electrical Engineering, from the University of Maiduguri (Nigeria) in 1991 and the University of Lagos, in 1998, respectively. He has been a lecturer at UNILAG since 2002. His research interests extend over antennas and propagation, Biomedical Instrumentation, and software development. Muhammed is a core member of the team that pioneered the use of the QMM for empirical modelling of engineering phenomena.

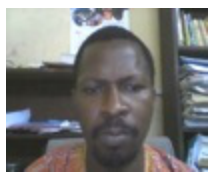

**Francis Okewole** was awarded the Bachelor of Technology (B. Tech.) Degree in Electrical Engineering, by the Ladoke Akintola University of Technology, Ogbomosho, Nigeria, in 2002, and obtained the degree of M. Sc. (also in Electrical Engineering) from the University of Lagos, in 2008. He is a Lecturer in the Department of Electrical and Electronics Engineering of the University of Lagos, and also a core member of the pioneering QMM modelling team. He has published a number of journal papers and conference proceedings papers in the area of antennas and radiowave propagation.

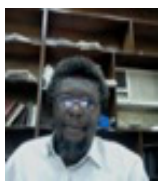

**Ike Mowete** obtained the degrees of B.Sc. (1980), M. Sc. (1983), and Ph.D. (1990) all in electrical engineering and from the University of Lagos, Akoka, Lagos, in Nigeria. His Ph.D. thesis was on a "quasi-static moment-method analysis of microstrip antennas". He has published quite a few papers on microstrip antennas, dielectric-costed thin-wire antennas, and shielding effectiveness of planar shields. His current research interests include thin-wire antenna structures, applications of the theory of characteristic modes to antenna analysis and design, and spectrum engineering issues. Professor Mowete, another core member of the QMM modeling pioneering team, teaches numerical methods, circuit theory and antennas and propagation at the University of Lagos, and is a member of the IEEE.